\documentclass[11pt]{article}

\usepackage{young}
\usepackage[latin1]{inputenc}
\usepackage[linesnumbered,boxed,ruled,commentsnumbered]{algorithm2e}
\usepackage{appendix}
\usepackage{epstopdf}
\usepackage{xcolor}
\usepackage{subcaption}
\usepackage[T1]{fontenc}
\usepackage[sc]{mathpazo}
\usepackage{amsmath}
\usepackage{amssymb}
\usepackage{graphicx,color}
\usepackage{framed}
\usepackage{multirow}
\usepackage{enumerate}
\usepackage{amsthm}
\usepackage{amsfonts,mathrsfs}
\usepackage{geometry} 
\usepackage{eepic}
\usepackage{ifthen}
\usepackage[vcentermath]{youngtab}
\usepackage[unicode=true,pdfusetitle, bookmarks=true,bookmarksnumbered=false,bookmarksopen=false, breaklinks=false,pdfborder={0 0 0},backref=false,colorlinks=false] {hyperref}
\hypersetup{
colorlinks,linkcolor=myurlcolor,citecolor=myurlcolor,urlcolor=myurlcolor}
\definecolor{myurlcolor}{rgb}{0,0,0.7}

\usepackage{color}

\newcommand{\blue}{\textcolor{blue}}

\usepackage{hyperref}
\hypersetup{pdfpagemode=UseNone}

\renewcommand{\det}{\operatorname{det}}

\def\I{\mathbb{1}}

\newenvironment{mylist}[1]{\begin{list}{}{
    \setlength{\leftmargin}{#1}
    \setlength{\rightmargin}{0mm}
    \setlength{\labelsep}{2mm}
    \setlength{\labelwidth}{8mm}
    \setlength{\itemsep}{0mm}}}
    {\end{list}}

\newcommand{\Pa}[1]{\left(#1\right)}

\newcommand{\bra}[1]{\langle#1|}

\newcommand{\ket}[1]{|#1\rangle}

\newcommand{\braket}[1]{\langle#1\rangle}

\DeclareMathOperator{\trace}{Tr}

\newcommand{\Ptr}[2]{\trace_{#1}\Pa{#2}}

\newcommand{\Tr}[1]{\Ptr{}{#1}}

\def\bsD{\boldsymbol{D}}
\def\bsG{\boldsymbol{G}}\def\bsH{\boldsymbol{H}}
\def\bsO{\boldsymbol{O}}
\def\bsP{\boldsymbol{P}}\def\bsQ{\boldsymbol{Q}}
\def\bsU{\boldsymbol{U}}\def\bsV{\boldsymbol{V}}\def\bsX{\boldsymbol{X}}

\theoremstyle{definition}

\numberwithin{equation}{section}

\newcounter{questionnumber}

\usepackage{graphicx}
\usepackage{subcaption}

\begin{document}

\title{\bf\Large Exact quantum algorithm for searching of multiple sets intersection}

\author{\blue{Sheng-Ao Mao}$^1$,\quad\blue{Lin Zhang}$^1$\footnote{E-mail: godyalin@163.com}, \quad \blue{Bo Li}$^2$\footnote{E-mail: libobeijing2008@163.com}\\
   {\it\small $^1$School of Mathematical Sciences, Hangzhou Dianzi University, Hangzhou 310018, PR~China}\\
{\it \small $^2$School of Computer and Computing Science, Hangzhou City University, Hangzhou 310015, China}}

\date{}
\maketitle

\begin{abstract}
Grover's algorithm achieves a quadratic speedup for unstructured search when a global oracle for the target set is available. However, its success probability is generally below $100\%$. To overcome this issue, Long presented a modified version of Grover's search algorithm by introducing a phase-matching condition, which can search for the target set with zero theoretical failure rate. Nevertheless, in many applications, the target set is determined by the intersection of multiple constrained sets. Most existing Grover-based methods are limited to the intersection of only two sets and are not always successful in searching for the target set. In this work, based on Long's method, we propose an algorithm for exactly searching the intersection of multiple sets. We derive and prove a phase-matching condition for searching the intersection exactly. Furthermore, we present a quantum circuit that implements exact intersection search. Finally, we support our theoretical results with numerical simulations.
\end{abstract}

\newpage

\section{Introduction}

Grover's algorithm is a fundamental quantum algorithm that achieves a quadratic speedup for unstructured search~\cite{grover1996fast}. However, its success probability is generally below $100\%$, which can sometimes make it less efficient than even guessing a solution, especially if there are particularly many. To overcome this issue, lots of modified versions of Grover's algorithm were proposed~\cite{grover1998quantum,long2001grover,hoyer2000arbitrary,roy2022deterministic,pokharel2211better}. Among them, the modified version proposed by Long~\cite{long2001grover} is the most representative work. A phase-matching condition was introduced to construct an exact search algorithm, and it has been widely used in the design of quantum search algorithms~\cite{li2007phase,toyama2008multiphase}. 

However, the standard formulation of these algorithms assumes a single oracle that defines the solution space. In many applications, the solution is specified implicitly as the intersection of multiple constraints. Such settings are frequently encountered in constraint satisfaction, combinatorial search~\cite{chen2022quantum,li2024resource,hogg2003adiabatic}, and database filtering tasks~\cite{salman2012quantum}. Moreover, constructing correct phase oracles for these constraints is itself a nontrivial task; recent work has begun to benchmark the ability of large language models to encode classical problems into verified quantum oracles~\cite{che2026qencodebench}. Related work on quantum set operations has been studied in~\cite{pang2013quantum,elgendy2024efficient,khadiev2023quantum}, but existing Grover-based methods are mostly restricted to the intersection of two sets. A recent randomized variant of Grover search method~\cite{dong2026random} has considered searching for the intersection of multiple sets, but the probability of finding the intersection is generally below $100\%$.

In this work, we propose an algorithm for exactly searching for the intersection of multiple sets based on the idea of Long's algorithm~\cite{long2001grover}. We derive and prove a phase-matching condition under which the search for the intersection succeeds with $100\%$ probability. Furthermore, we present a quantum circuit for exact intersection search and use numerical simulations to support our theoretical results.

The remainder of this paper is organized as follows. In Sec.~\ref{sec:2}, we briefly review the original Grover's algorithm and its modified version, and we state the problem. In Sec.~\ref{sec:3}, we present a quantum algorithm for exactly searching for the intersection of multiple sets. In particular, we derive and prove the phase-matching condition in this section. The quantum circuit construction and numerical simulations are presented in Sec.~\ref{sec:4}. Finally, Sec.~\ref{sec:5} concludes the paper.

\section{Preliminaries}
\label{sec:2}
\subsection{The original Grover's algorithm}
Grover's algorithm is a quantum search procedure for finding a target element in an unstructured database of size $N=2^n$, providing a quadratic speedup over classical search methods~\cite{grover1996fast}. Let $\mathcal{M}\subseteq\{0,\ldots,N-1\}$ denote the set of target elements, with $|\mathcal{M}|=r$. The algorithm begins with the uniform superposition
\[
\ket{\psi}
=
\frac{1}{\sqrt{N}}
\sum_{x=0}^{N-1}\ket{x},
\]
and repeatedly applies the Grover operator $\bsG=\bsD\bsO$, where $\bsO$ is the phase oracle that marks the elements in $\mathcal{M}$ and $\bsD=2\ket{\psi}\bra{\psi}-\I$ is the diffusion operator.
The evolution of Grover's algorithm is confined to the two-dimensional subspace spanned by the normalized superpositions of the target and non-target states,
\[
\ket{\mu}
=
\frac{1}{\sqrt{r}}
\sum_{x\in\mathcal{M}}\ket{x},
\qquad
\ket{\nu}
=
\frac{1}{\sqrt{N-r}}
\sum_{x\notin\mathcal{M}}\ket{x}.
\]
In this basis, the Grover operator acts as a rotation by a fixed angle determined by $\sin\theta=\sqrt{\frac{r}{N}}.$
Therefore, after $m$ Grover iterations, the state is given by
\[
\ket{\psi_m}
=
\sin\bigl((2m+1)\theta\bigr)\ket{\mu}
+
\cos\bigl((2m+1)\theta\bigr)\ket{\nu}.
\]
Thus, the probability of measuring an element in $\mathcal{M}$ is
\[
P(m)
=
\sin^2\bigl((2m+1)\theta\bigr).
\]
The probability approaches 1 as $(2m+1)\theta$ approaches $\pi/2$. Accordingly, the optimal number of Grover iterations is approximately $m\approx\frac{\pi}{4\theta}-\frac{1}{2}.$
Since the number of iterations must be an integer, $(2m+1)\theta$ generally cannot be exactly equal to $\pi/2$, and hence the maximum probability is not necessarily equal to 1. Actually, it has been rigorously proven that the original Grover's algorithm is exact only when searching for one out of four~\cite{diao2010exactness}.

\subsection{The modified Grover's algorithm}
To solve the shortcomings mentioned above, Long proposed an exact quantum search algorithm based on a phase-matching condition~\cite{long2001grover}. Instead of using the $\pi$-phase inversions in the original Grover operator, the modified algorithm replaces them with identical phase rotations of an appropriate angle.
The modified Grover's algorithm begins with the uniform superposition
\[
\ket{\psi}
=
\frac{1}{\sqrt{N}}
\sum_{x=0}^{N-1}\ket{x}.
\]
The modified oracle and diffusion operators are defined as
\begin{eqnarray*}
\bsO_{L}=\I+\left(e^{i\phi}-1\right)\sum_{x\in\mathcal{M}}\ket{x}\bra{x},\qquad\bsD_{L}=\I+\left(e^{i\phi}-1\right)\ket{\psi}\bra{\psi},
\end{eqnarray*}
respectively. The corresponding iteration operator is $\bsQ=-\bsD_{L}\bsO_{L}.$
To satisfy the phase-matching condition, the two phase-rotation angles are chosen to be equal and given by
\[
\phi
=
2\arcsin
\left(
\tfrac{
	\sin\left(\tfrac{\pi}{4J+6}\right)
}{
	\sin\theta
}
\right),
\]
where $J\geqslant J_{\min}:=\left\lfloor\frac{\frac{\pi}{2}-\theta}{2\theta}\right\rfloor$, the integer part of $\frac{\frac{\pi}{2}-\theta}{2\theta}$. Here, $J$ can be any integer satisfying $J\geqslant J_{\min}$. By applying the operator $\bsQ$ for $J+1$ iterations, the probability of measuring an element in $\mathcal{M}$ becomes 1. 

\subsection{Problem statement}
We consider a search space $\mathcal{S}$ of size $N = 2^n$, where $n$ is a positive integer. Let $\mathcal{A}_0, \mathcal{A}_1, \ldots, \mathcal{A}_{k-1} \subset \mathcal{S}$ be $k$ distinct nonempty subsets, and let $\bsG_0, \bsG_1, \ldots, \bsG_{k-1}$ denote the corresponding Grover operators. We define the intersection of these subsets as
\[
\mathcal{A} := \bigcap_{i=0}^{k-1} \mathcal{A}_i,
\]
where $0 < |\mathcal{A}| = r < N$. Note that \( \bsG_i = \bsD\bsO_i \), where
\[
\bsD = 2 |\psi\rangle\langle\psi| - \I, \quad \text{and} \quad 
\bsO_i|x\rangle =
\begin{cases} 
	-|x\rangle, & x \in \mathcal{A}_i, \\
	|x\rangle, & x \notin \mathcal{A}_i.
\end{cases}
\]
The goal of this work is to construct a global oracle for $\mathcal{A}$ using the oracles of the subsets, enabling exact quantum search for the intersection.
\section{Exact quantum algorithm for searching of multiple sets intersection}
\label{sec:3}
Based on the idea of the modified Grover's algorithm, we construct an operator that exactly finds the intersection of multiple sets. The construction method and its proof are given below.

To construct an oracle that directly identifies the intersection
$\mathcal{A}$, define
\begin{equation}
	\Pi_i=\frac{\I-\bsO_i}{2}.
	\label{eq:Pi_i}
\end{equation}
Then
\[
\Pi_i\ket{x}
=
\begin{cases}
	\ket{x}, & x\in\mathcal{A}_i,\\
	0,       & x\notin\mathcal{A}_i.
\end{cases}
\]
Thus, $\Pi_i$ projects onto the subspace associated with $\mathcal{A}_i$.
Since all $\bsO_i$ are diagonal in the computational basis, the projectors
$\Pi_i$ commute. Hence, the projector onto the intersection is
\begin{equation}
	\Pi
	=
	\prod_{i=0}^{k-1}\Pi_i
	=
	\prod_{i=0}^{k-1}\frac{\I-\bsO_i}{2}.
	\label{eq:intersection_projector}
\end{equation}
It satisfies
\begin{equation}
	\Pi\ket{x}
	=
	\begin{cases}
		\ket{x}, & x\in\mathcal{A},\\
		0,       & x\notin\mathcal{A}.
	\end{cases}
	\label{eq:intersection_projector_action}
\end{equation}
Based on $\Pi$, we define the phase oracle associated with the
intersection as
\begin{equation}
	\bsO_{\cap}(\phi)
	=
	\I+\left(e^{i\phi}-1\right)\Pi.
	\label{eq:intersection_phase_oracle}
\end{equation}
By Eq.~\eqref{eq:intersection_projector_action}, we obtain
\begin{equation}
	\bsO_{\cap}(\phi)\ket{x}
	=
	\begin{cases}
		e^{i\phi}\ket{x}, & x\in\mathcal{A},\\
		\ket{x},          & x\notin\mathcal{A}.
	\end{cases}
	\label{eq:intersection_oracle_action}
\end{equation}
Hence, $\bsO_{\cap}(\phi)$ applies the same phase rotation $e^{i\phi}$ to all states in the intersection while leaving all other computational basis states unchanged. Next, we define the diffusion operator as
\begin{equation}
	\bsD(\phi)
	=
	-\left[
	\I+\left(e^{i\phi}-1\right)
	\ket{\psi}\bra{\psi}
	\right].
\end{equation}
We can then construct the intersection search operator as
\begin{equation}
	\bsG_{\cap}(\phi)
	=
	\bsD(\phi)\bsO_{\cap}(\phi),
	\label{eq:exact_search_operator}
\end{equation}
where the phase $\phi$ and the iteration number $L$ satisfy
\begin{align}
	\label{eq:phase-matching condition}
	\phi
	&=2\arcsin\left(\tfrac{\sin\left(\tfrac{\pi}{4L+2}\right)}{\sin\beta}\right), \\
	L
	&\geqslant
	L_{\min}
	=
	\left\lfloor
	\tfrac{\tfrac{\pi}{2}+\beta}{2\beta}
	\right\rfloor, \\
	\beta
	&=\arcsin\sqrt{\tfrac{r}{N}}.
\end{align}
Here the two-phase rotations are equal which is required by the phase-matching condition. Exact search for the intersection is achieved by measuring the quantum computer after $L$ iterations.

In the following part, we provide the proof of the above result. First, we define the normalized superpositions of the target and non-target states by
\begin{equation}
	\ket{T}
	=
	\frac{1}{\sqrt{r}}
	\sum_{x\in\mathcal{A}}\ket{x},
	\qquad
	\ket{R}
	=
	\frac{1}{\sqrt{N-r}}
	\sum_{x\notin\mathcal{A}}\ket{x}.
	\label{eq:TR_states}
\end{equation}
Then the initial state can be expressed as
\begin{equation}
	\ket{\psi}
	=
	\sin\beta\,\ket{T}
	+
	\cos\beta\,\ket{R},
	\label{eq:psi_TR}
\end{equation}
where
\begin{equation}
	\sin\beta=\sqrt{\tfrac{r}{N}},
	\qquad
	\cos\beta=\sqrt{1-\tfrac{r}{N}}.
	\label{eq:beta_definition}
\end{equation}
Both $\bsO_{\cap}(\phi)$ and $\bsD(\phi)$ preserve the two-dimensional
subspace spanned by $\{\ket{T},\ket{R}\}$. Therefore, the entire
evolution can be analyzed within this subspace.

In the basis $\{\ket{T},\ket{R}\}$, the diffusion operator is
represented by
\begin{equation}
	\bsD(\phi)
	=
	-
	\begin{pmatrix}
		1+(e^{i\phi}-1)\sin^2\beta
		&
		(e^{i\phi}-1)\sin\beta\cos\beta
		\\
		(e^{i\phi}-1)\sin\beta\cos\beta
		&
		1+(e^{i\phi}-1)\cos^2\beta
	\end{pmatrix},
	\label{eq:D_matrix}
\end{equation}
whereas the intersection oracle takes the form
\begin{equation}
	\bsO_{\cap}(\phi)
	=
	\begin{pmatrix}
		e^{i\phi} & 0\\
		0 & 1
	\end{pmatrix}.
	\label{eq:Ocap_matrix}
\end{equation}
Consequently,
\begin{equation}
	\bsG_{\cap}(\phi)
	=
	-
	\begin{pmatrix}
		e^{i\phi}
		\left[1+(e^{i\phi}-1)\sin^2\beta\right]
		&
		(e^{i\phi}-1)\sin\beta\cos\beta
		\\
		e^{i\phi}(e^{i\phi}-1)\sin\beta\cos\beta
		&
		1+(e^{i\phi}-1)\cos^2\beta
	\end{pmatrix}.
	\label{eq:Gcap_matrix}
\end{equation}
Since a global phase has no effect on measurement probabilities, we set
\begin{eqnarray}	\label{eq:V_definition}
\bsV= -e^{-i\phi}\bsG_{\cap}(\phi) = 
		\begin{pmatrix}
			\cos^2\beta+e^{i\phi}\sin^2\beta
			&
			(1-e^{-i\phi})\sin\beta\cos\beta
			\\
			(e^{i\phi}-1)\sin\beta\cos\beta
			&
			\cos^2\beta+e^{-i\phi}\sin^2\beta
		\end{pmatrix}.
\end{eqnarray}
A direct calculation gives $\det \bsV=1$. Moreover, since
$\bsG_{\cap}(\phi)$ is unitary, $\bsV$ is also unitary, and therefore $\bsV\in SU(2).$
The trace of $\bsV$ is
\begin{eqnarray}\label{eq:trace_V}
\Tr{\bsV}=2\cos^2\beta+2\cos\phi\,\sin^2\beta=2-4\sin^2\beta\sin^2\frac{\phi}{2}=
2\cos(2\delta),
\end{eqnarray}
where $\sin\delta=\sin\beta\sin\frac{\phi}{2}$. Therefore, the characteristic polynomial of $\bsV$ is
\begin{equation}
	\lambda^2
	-
	2\cos(2\delta)\lambda
	+
	1
	=
	0,
\end{equation}
and its eigenvalues are $\lambda_{\pm}=e^{\pm i2\delta}$. By the Cayley--Hamilton theorem,
\begin{equation}
	\bsV^2
	-
	2\cos(2\delta)\bsV
	+
	\I
	=
	0.
	\label{eq:CH_V}
\end{equation}
This implies that for $L \geqslant 2$,
\begin{equation}
	\bsV^L
	=
	\frac{\sin(2L\delta)}{\sin(2\delta)}\bsV
	-
	\frac{\sin(2(L-1)\delta)}{\sin(2\delta)}\I.
	\label{eq:V_power}
\end{equation}
After $L$ iterations, the non-target amplitude is given by
\begin{equation}
	B_L
	=
	\bra{R}\bsV^L\ket{\psi}.
	\label{eq:BL_definition}
\end{equation}
From Eqs.~\eqref{eq:V_definition} and \eqref{eq:psi_TR}, we obtain
\begin{equation}
	\bra{R}\bsV\ket{\psi}
	=
	\left(1-4\sin^2\delta\right)\cos\beta,
	\qquad
	\braket{R|\psi}
	=
	\cos\beta.
	\label{eq:R_V_psi}
\end{equation}
Substituting these expressions into Eq.~\eqref{eq:V_power} gives
\begin{align}
	B_L
	&=
	\frac{\cos\beta}{\sin(2\delta)}
	\Big[
	(1-4\sin^2\delta)\sin(2L\delta)
	-
	\sin(2(L-1)\delta)
	\Big]\nonumber\\
	&=
	\frac{\cos\beta}{\cos\delta}
	\cos\bigl((2L+1)\delta\bigr).
	\label{eq:BL_intermediate}
\end{align}
Therefore, exact search for the intersection is achieved if and only if $B_L=0$, which requires $\cos\bigl((2L+1)\delta\bigr)=0$, that is, $\delta=\frac{\pi}{4L+2}$. Since $\sin\delta=\sin\beta\sin\frac{\phi}{2}$, we conclude that exact search for the intersection is achieved when
\[
\phi = 2\arcsin\left(\tfrac{\sin\left(\tfrac{\pi}{4L+2}\right)}{\sin\beta}\right).
\]
Eq.~\eqref{eq:phase-matching condition} has real solutions for $L \geqslant L_{\min}$, which is the minimum in most cases.

\section{Quantum circuit implementation and numerical simulations}
\label{sec:4}
In this section, we present the quantum circuit construction for exact intersection search of multiple sets, and we provide numerical experiments for $k=2$, $3$, and $4$. 

\subsection{Quantum circuit implementation for the exact intersection search}
\label{subsec:exact_intersection_circuit}
\begin{figure}[htbp]
	\centering
	\includegraphics[width=\linewidth]{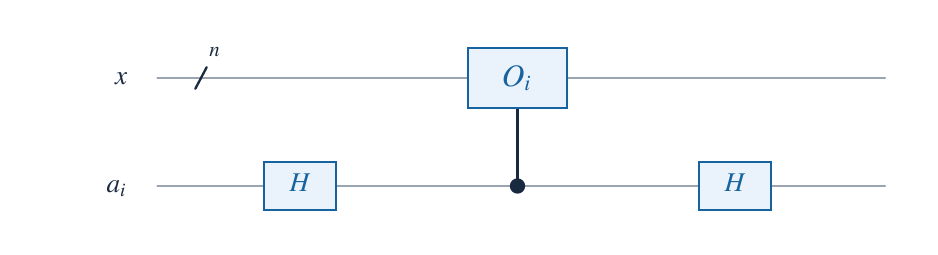}
	\caption{Quantum circuit for the membership oracle $\bsU_{f_i}$.}
	\label{fig:41_membership_oracle_circuit}
\end{figure}

We now present the quantum circuit implementations of the operators introduced in Sec.~\ref{sec:3}. For each set $\mathcal{A}_i$, define its membership function by
\begin{equation}
	f_i(x)=
	\begin{cases}
		1, & x\in\mathcal{A}_i,\\
		0, & x\notin\mathcal{A}_i.
	\end{cases}
	\label{eq:41_membership}
\end{equation}
Therefore, $\bsO_i$ can be written as $\bsO_i|x\rangle=(-1)^{f_i(x)}|x\rangle$, so that $\bsO_{\cap}(\phi)|x\rangle=e^{i\phi\prod_{i=0}^{k-1}f_i(x)}|x\rangle$. To compute $f_i(x)$ coherently, we assume that the controlled phase oracle is accessible and thus define
\begin{equation}
	\bsU_{f_i}
	=
	(\I_x\otimes \bsH_{a_i})\,
	\mathrm{C}_{a_i \rightarrow x} \bsO_i
	(\I_x\otimes \bsH_{a_i}),
	\label{eq:controlled_to_membership}
\end{equation}
where $a_i$ is the auxiliary qubit that stores the membership value and $\bsH_{a_i}$ is a Hadamard gate on that qubit. The quantum circuit implementing \(\bsU_{f_i}\) is shown in Fig.~\ref{fig:41_membership_oracle_circuit}.

\begin{figure}[htbp]
	\centering
	\includegraphics[width=\linewidth]{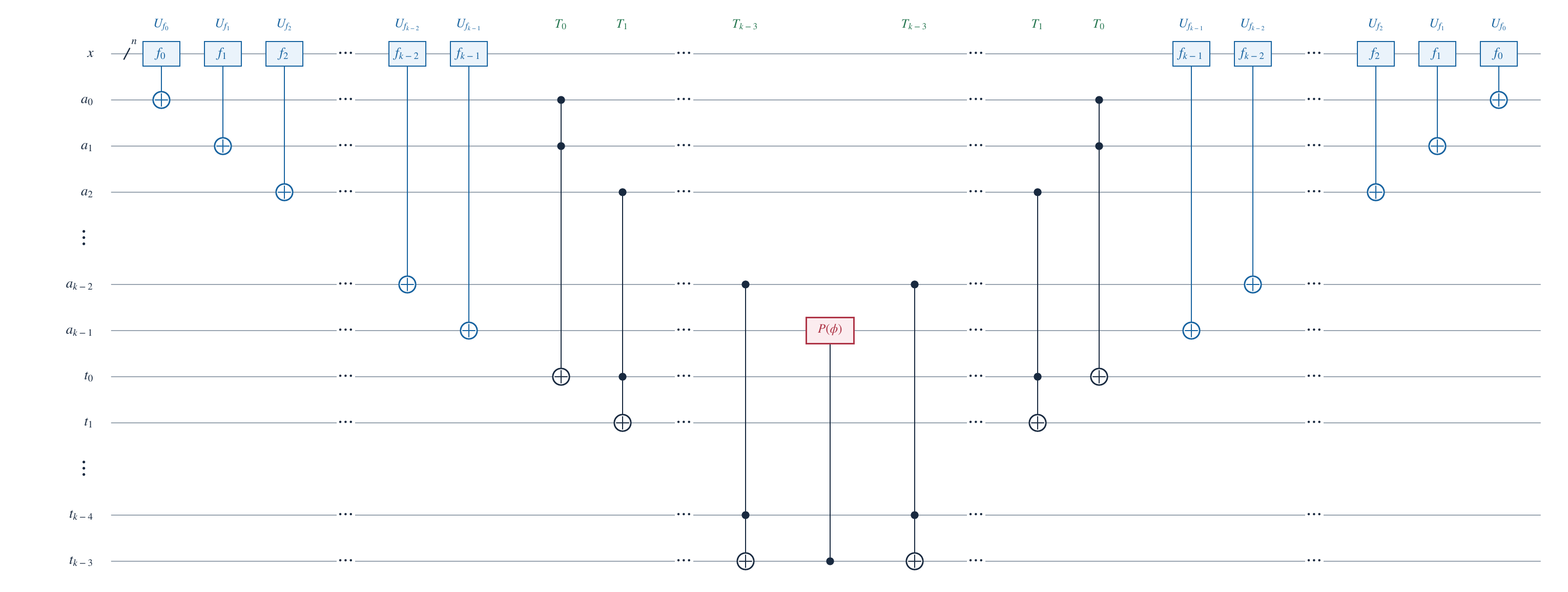}
	\caption{Quantum circuit for the intersection oracle operator $\bsO_{\cap}(\phi)$.}
	\label{fig:41_intersection_oracle}
\end{figure}

For $k\geqslant3$, we introduce $k$ auxiliary qubits $a_0,\ldots,a_{k-1}$ and $k-2$ auxiliary qubits $t_0,\ldots,t_{k-3}$, all initialized in $|0\rangle$. Define $\mathcal{U}=\bsU_{f_{k-1}}\cdots \bsU_{f_1}\bsU_{f_0}$, with each $\bsU_{f_i}$ acting on the qubits $x$ and the auxiliary qubit $a_i$. Let
\begin{align}
	T_0&=\mathrm{CCX}_{a_0,a_1\, \rightarrow \,t_0},\nonumber\\
	T_j&=\mathrm{CCX}_{t_{j-1},a_{j+1}\, \rightarrow \,t_j},
	\quad 1\leqslant j\leqslant k-3,\nonumber\\
	\mathcal{T}&=T_{k-3}\cdots T_1T_0,
	\label{eq:41_and_chain}
\end{align}
where CCX denotes the Toffoli gate. Writing $\bsP(\phi)=\operatorname{diag}(1,e^{i\phi})$, the circuit implementation of the intersection oracle is
\begin{equation}
	\bsO_{\cap}(\phi)
	=\mathcal{U}^{\dagger}\mathcal{T}^{\dagger}
	\mathrm{CP}_{t_{k-3}\rightarrow a_{k-1}}(\phi)
	\mathcal{T}\mathcal{U},
	\qquad k\geqslant3.
	\label{eq:41_oracle_circuit}
\end{equation}
For $k=2$, this expression becomes
$\bsO_{\cap}(\phi)=\mathcal{U}^{\dagger}\mathrm{CP}_{a_0\rightarrow a_1}(\phi)\mathcal{U}$.
Therefore, for \(k \geqslant 2\), we have
\begin{align}
	\bsO_{\cap}(\phi)|x\rangle|0\rangle_{\mathrm{aux}}
	&=e^{i\phi\prod_{i=0}^{k-1}f_i(x)}
	|x\rangle|0\rangle_{\mathrm{aux}}.
	\label{eq:41_oracle_equivalence}
\end{align}
The quantum circuit implementation of Eq.~\eqref{eq:41_oracle_circuit} is shown in Fig.~\ref{fig:41_intersection_oracle}.

\begin{figure}[htbp]
	\centering
	\includegraphics[width=0.72\linewidth]{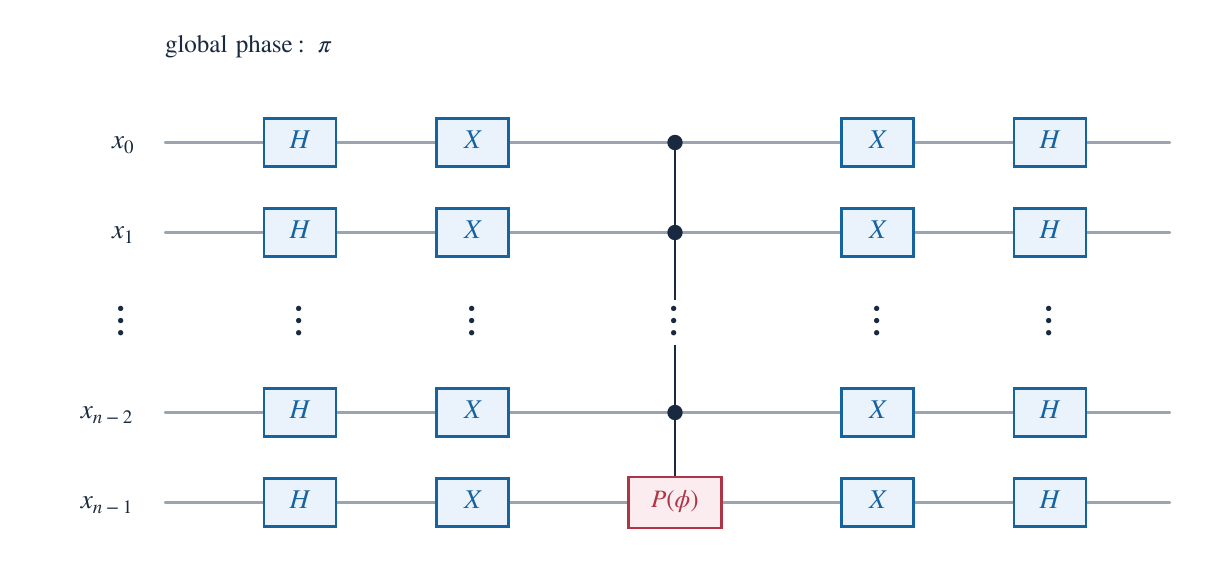}
	\caption{Quantum circuit for the diffusion operator $\bsD(\phi)$.}
	\label{fig:41_diffusion}
\end{figure}

The diffusion operator can be decomposed as
\begin{align}
	\bsD(\phi)
	&=-\left[\I+(e^{i\phi}-1)
	|\psi\rangle\langle\psi|\right]\nonumber\\
	\label{eq:41_diffusion_circuit}
	&=-\bsH^{\otimes n}\bsX^{\otimes n}
	\bigl[\mathrm{C}^{n-1}P(\phi)\bigr]
	\bsX^{\otimes n}\bsH^{\otimes n},
\end{align}
where
\begin{equation}
	\mathrm{C}^{n-1}\bsP(\phi)
	=\I+(e^{i\phi}-1)|1^n\rangle\langle1^n|.
	\label{eq:41_multicontrolled_phase}
\end{equation}
Fig.~\ref{fig:41_diffusion} shows the circuit corresponding to Eq.~\eqref{eq:41_diffusion_circuit}. Here, $x_0,x_1,\ldots,x_{n-1}$ denote the $n$ data qubits encoding the computational basis state $\ket{x}$. The $\bsX$ gates denote Pauli-$\bsX$ operations.

\begin{figure}[htbp]
	\centering
	\includegraphics[width=\linewidth]{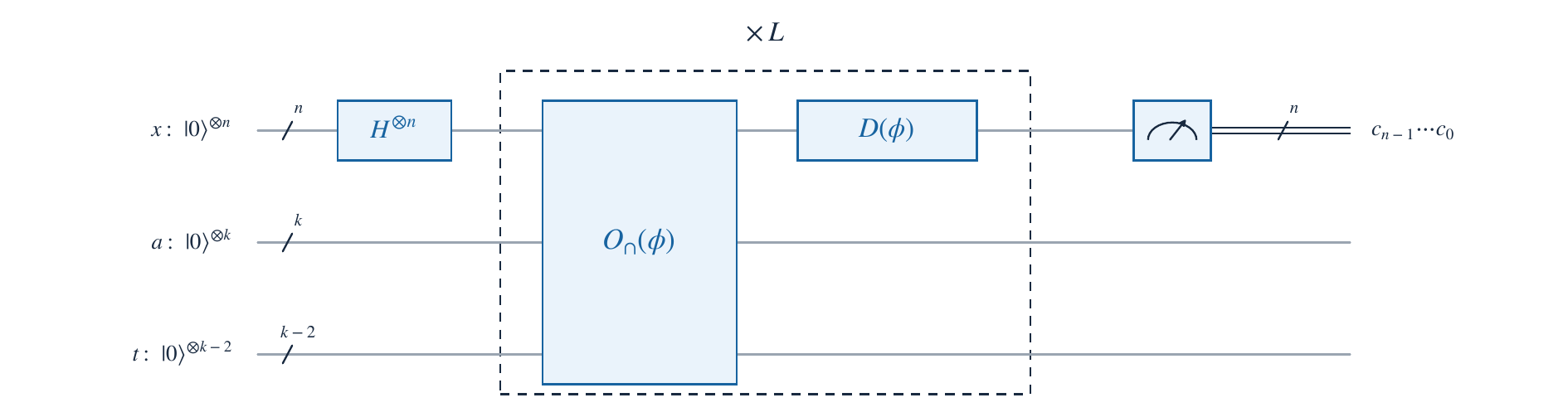}
	\caption{Complete quantum circuit for exact intersection search.}
	\label{fig:41_complete_search}
\end{figure}

Combining the two quantum circuits above, we present the complete quantum circuit implementation for exact intersection search, as shown in Fig.~\ref{fig:41_complete_search}. Note that implementing this circuit requires $2k-2$ auxiliary qubits.

\subsection{Numerical simulations}
In this subsection, we set the search space to $N=2^{5}$ and the number of quantum circuit shots to 1000. The number of iterations is set to $L=L_{\min}$, and $\phi$ is obtained from the phase-matching condition. We then consider the cases $k=2$, $3$, and $4$.
 
\begin{figure*}[htbp]
 	\centering
 	\includegraphics[width=0.49\textwidth]
 	{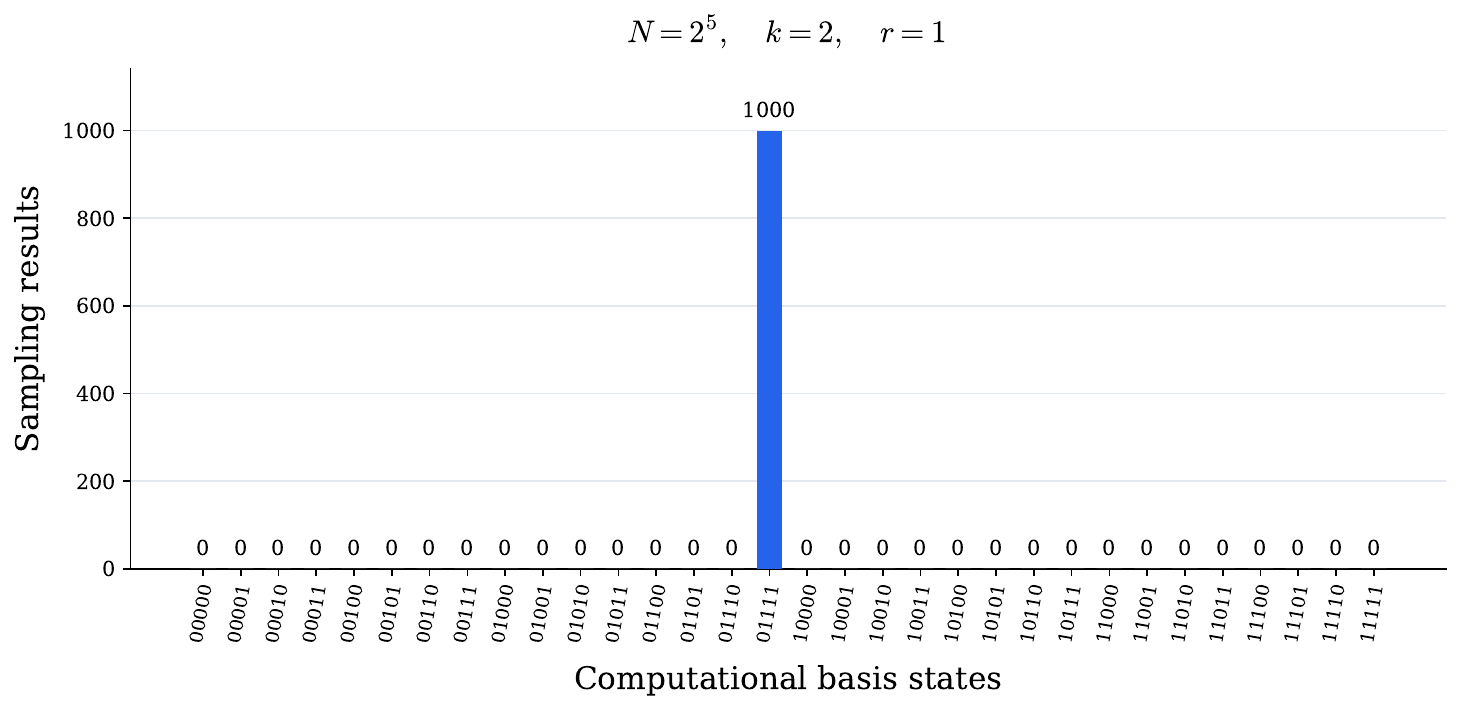}
 	\includegraphics[width=0.49\textwidth]
 	{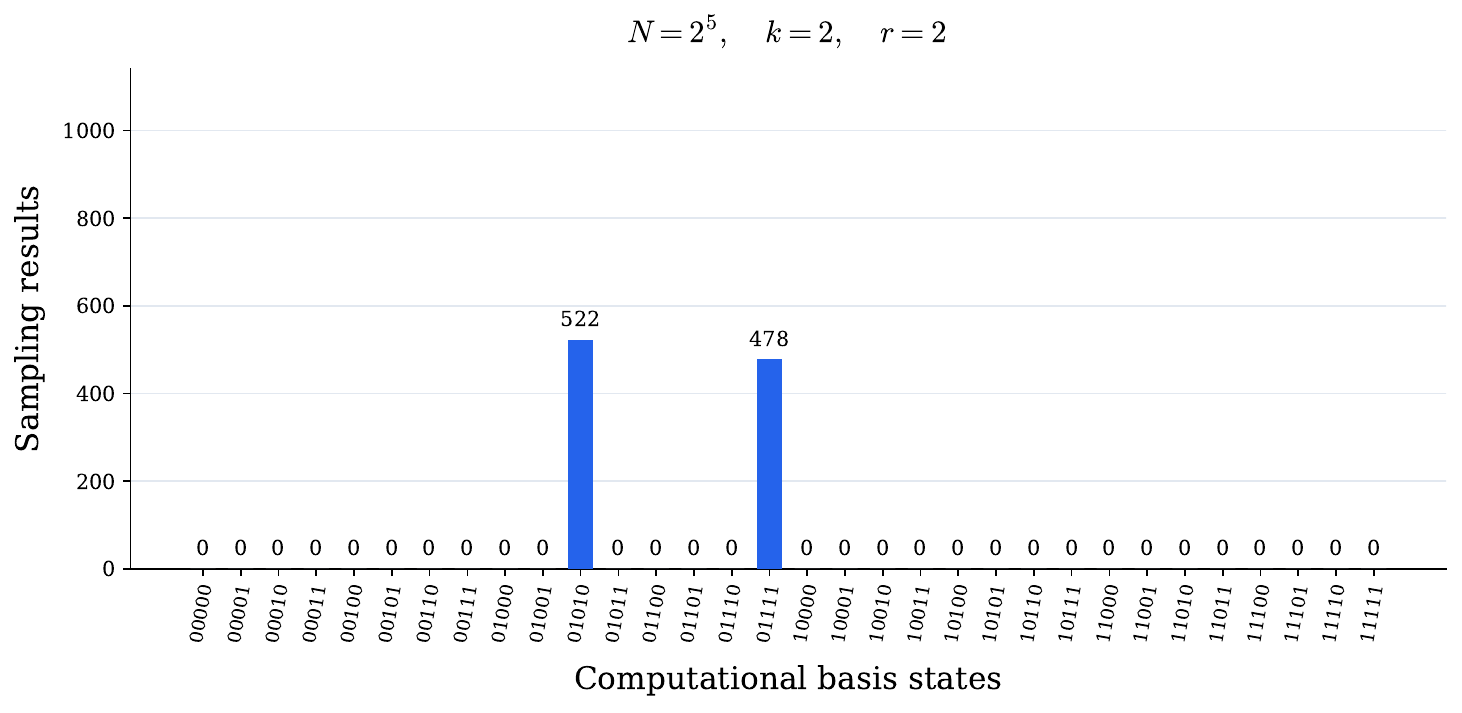}
 	\includegraphics[width=0.49\textwidth]
 	{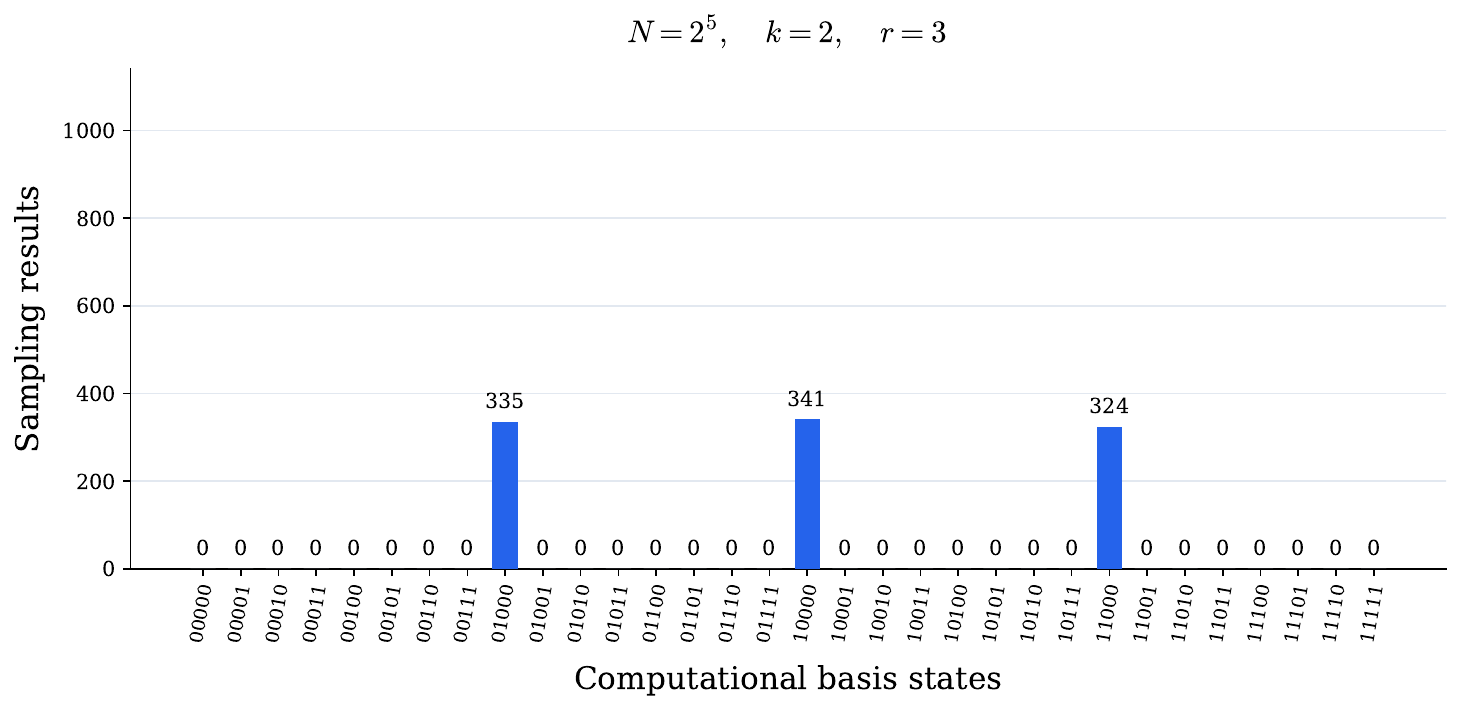}
 	\includegraphics[width=0.49\textwidth]
 	{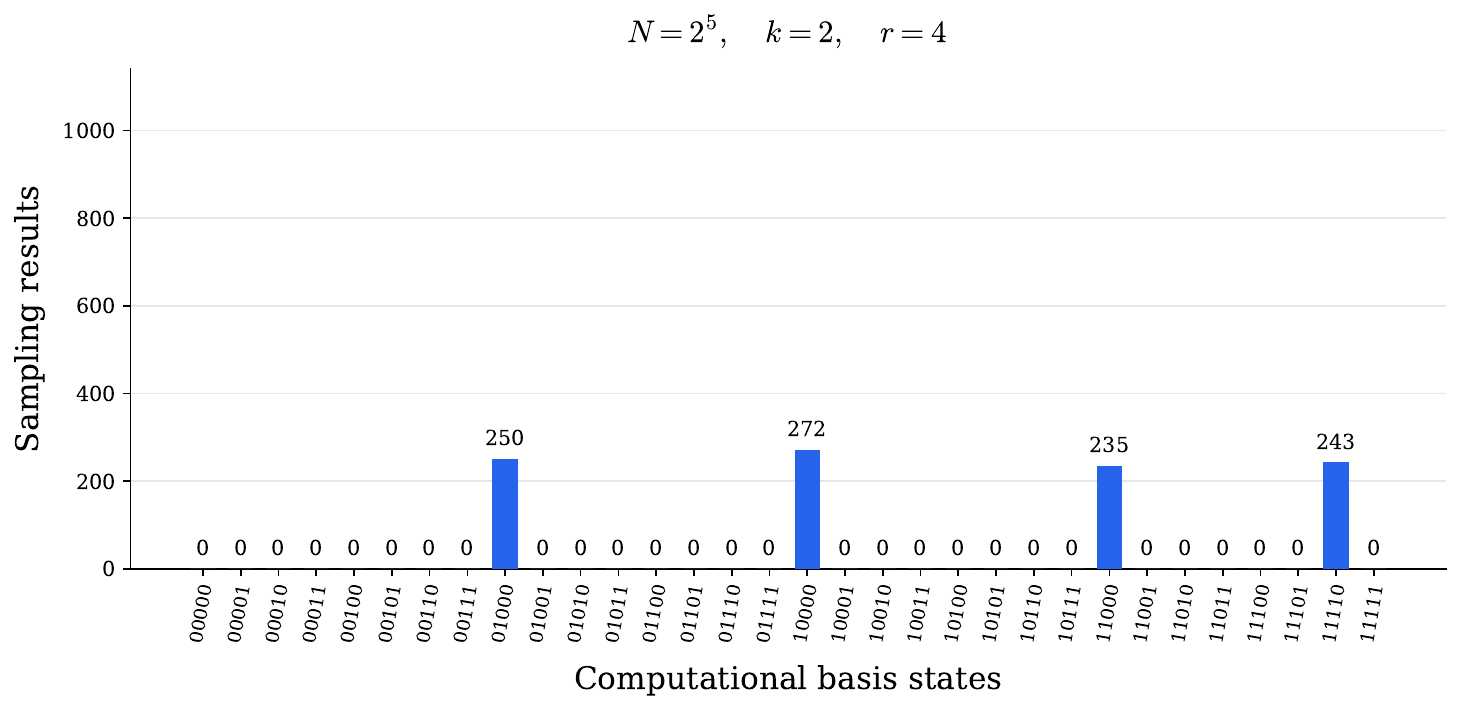}
 	\caption{
 		Experimental results for the exact search of the intersection of two sets for $r=1$, $2$, $3$, and $4$.
 	}
 	\label{fig:k2-results}
\end{figure*}

Fig.~\ref{fig:k2-results} shows the sampling results for $k=2$ with $r=1, 2, 3,$ and $4$. For $r=1, 2, 3,$ and $4$, the target states correspond to $\{|01111\rangle\}$, $\{|01010\rangle, |01111\rangle\}$, $\{|01000\rangle, |10000\rangle, |11000\rangle\}$, and $\{|01000\rangle, |10000\rangle, |11000\rangle, |11110\rangle\}$, respectively. The numbers of iterations $L$ are 4, 3, 3, and 2, respectively. In Fig.~\ref{fig:k2-results}, by sampling the quantum circuit 1000 times, the total number of detections of the target states in each case equals the number of quantum circuit shots, whereas no non-target states are detected. This result indicates that the intersection is searched with probability 100\%.

\begin{figure*}[htbp]
	\centering
 	\includegraphics[width=0.49\textwidth]
 	{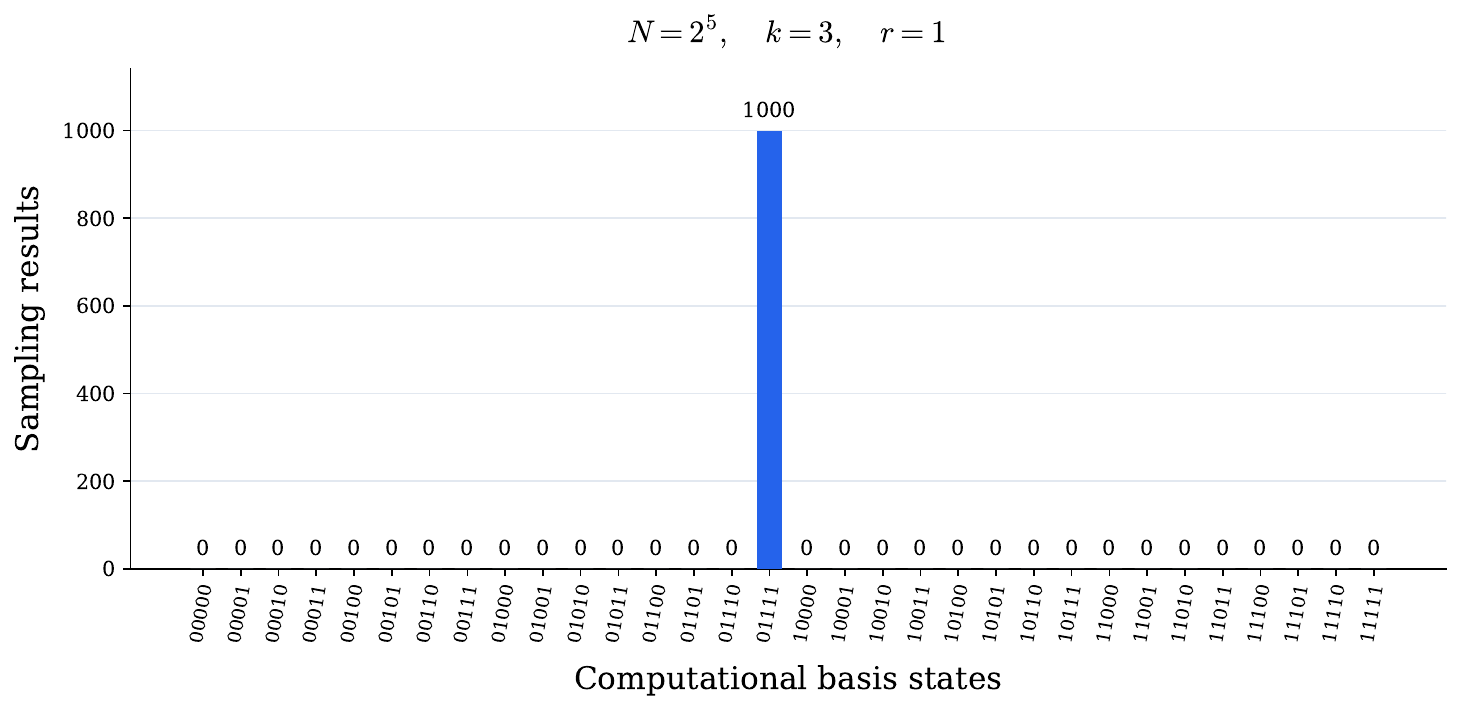}
	\includegraphics[width=0.49\textwidth]
 	{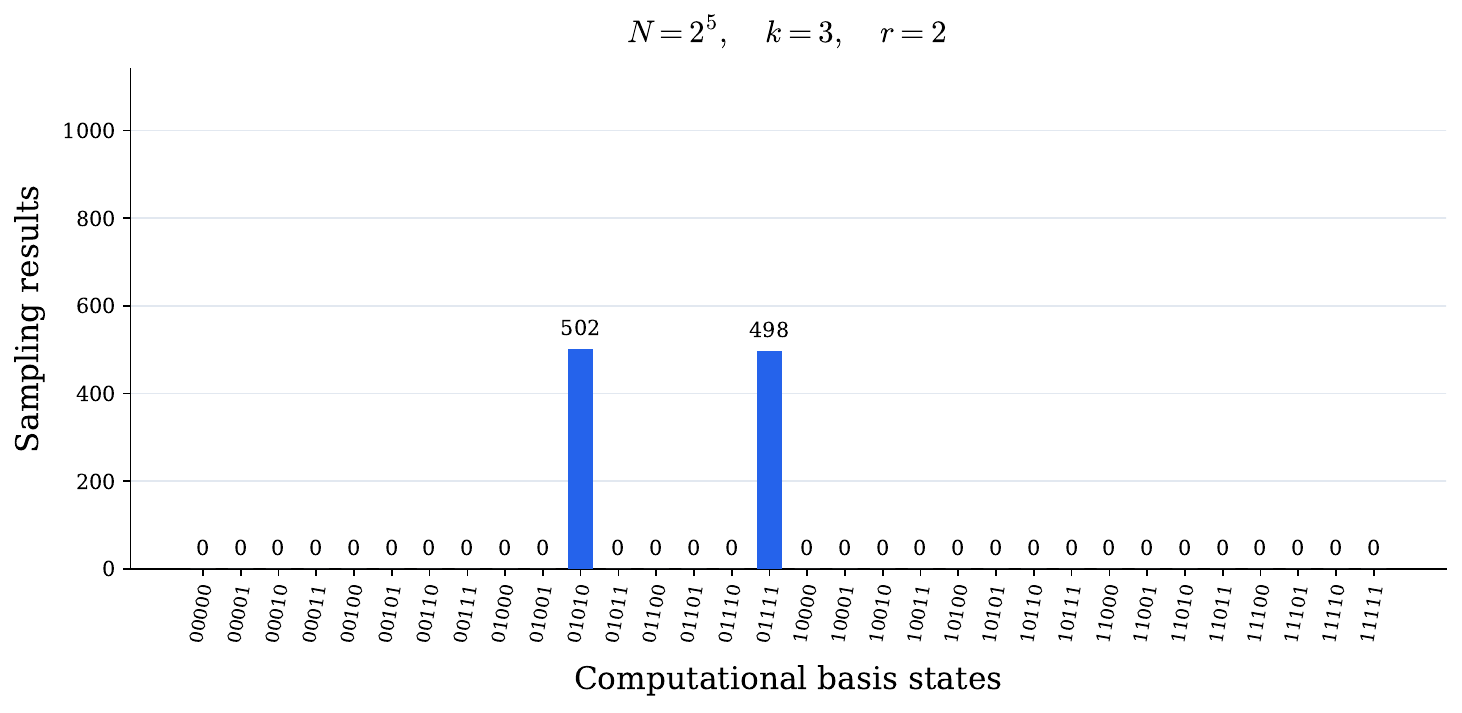}
 	\includegraphics[width=0.49\textwidth]
 	{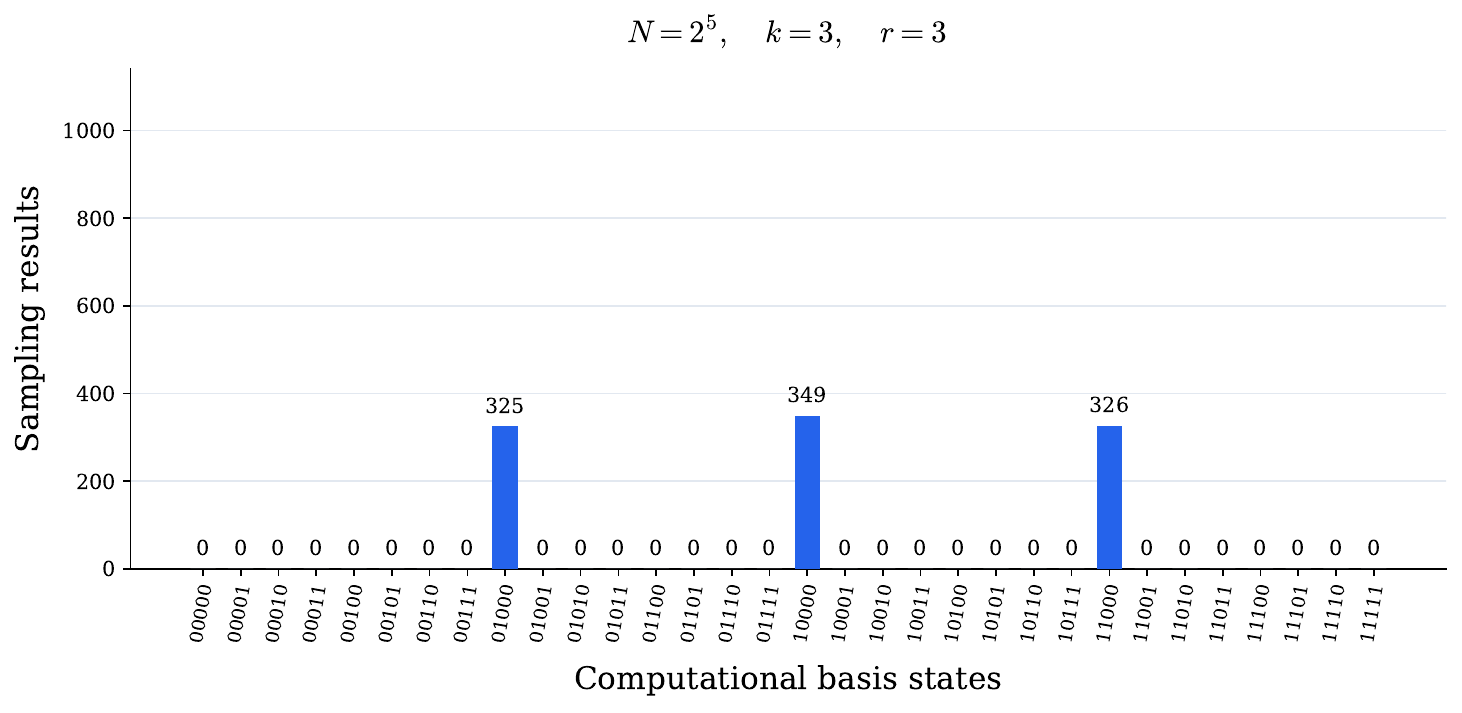}
 	\includegraphics[width=0.49\textwidth]
 	{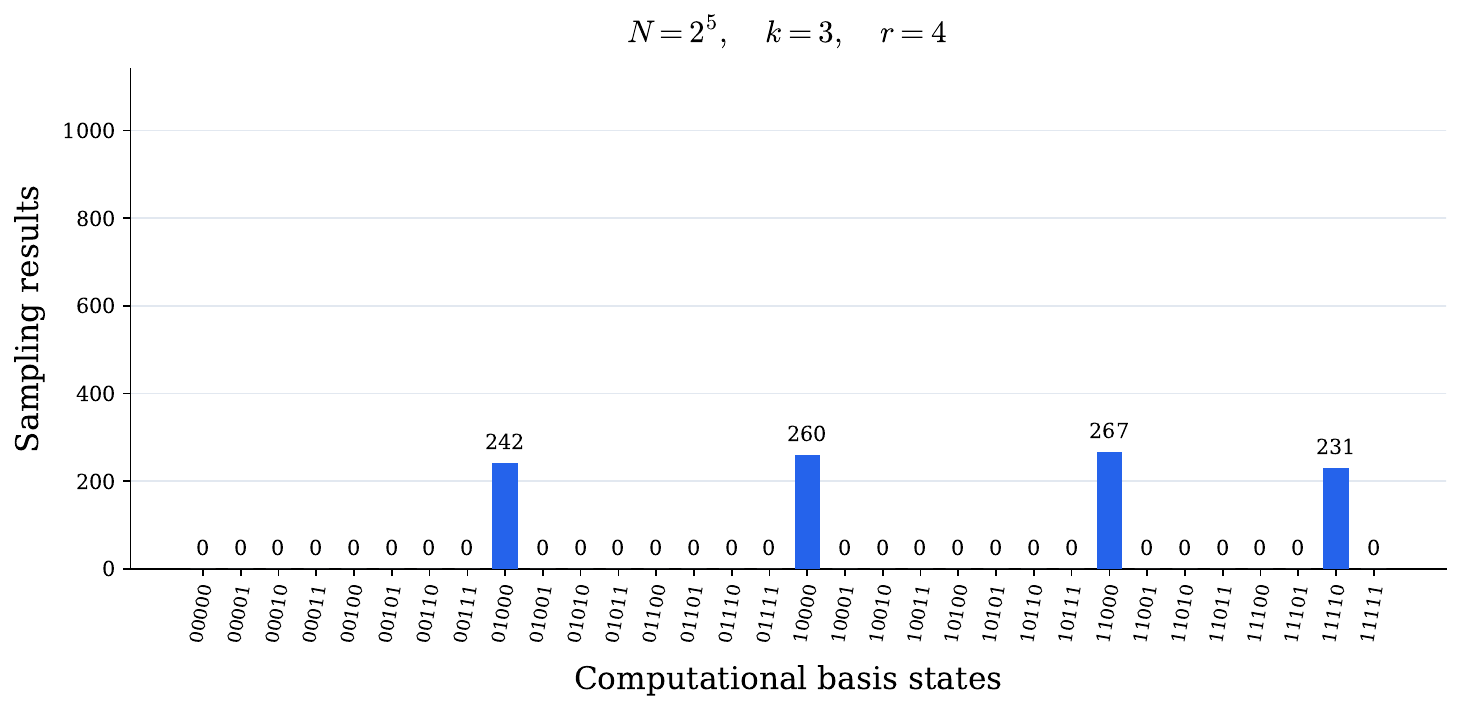}
 	\caption{
 		Experimental results for the exact search of the intersection of three sets for $r=1$, $2$, $3$, and $4$.
 	}
 	\label{fig:k3-results}
\end{figure*}

Fig.~\ref{fig:k3-results} shows the sampling results for $k=3$ with $r=1, 2, 3,$ and $4$. For $r=1, 2, 3,$ and $4$, the target states correspond to $\{|01111\rangle\}$, $\{|01010\rangle, |01111\rangle\}$, $\{|01000\rangle, |10000\rangle, |11000\rangle\}$, and $\{|01000\rangle, |10000\rangle, |11000\rangle, |11110\rangle\}$, respectively. The numbers of iterations $L$ are 4, 3, 3, and 2, respectively. From Fig.~\ref{fig:k3-results}, we observe that, with 1000 shots of the quantum circuit, the total number of samples for the target states in each case is 1000. This indicates that the intersection can be searched for with probability 100\%.

\begin{figure*}[htbp]
 	\centering
 	\includegraphics[width=0.49\textwidth]
 	{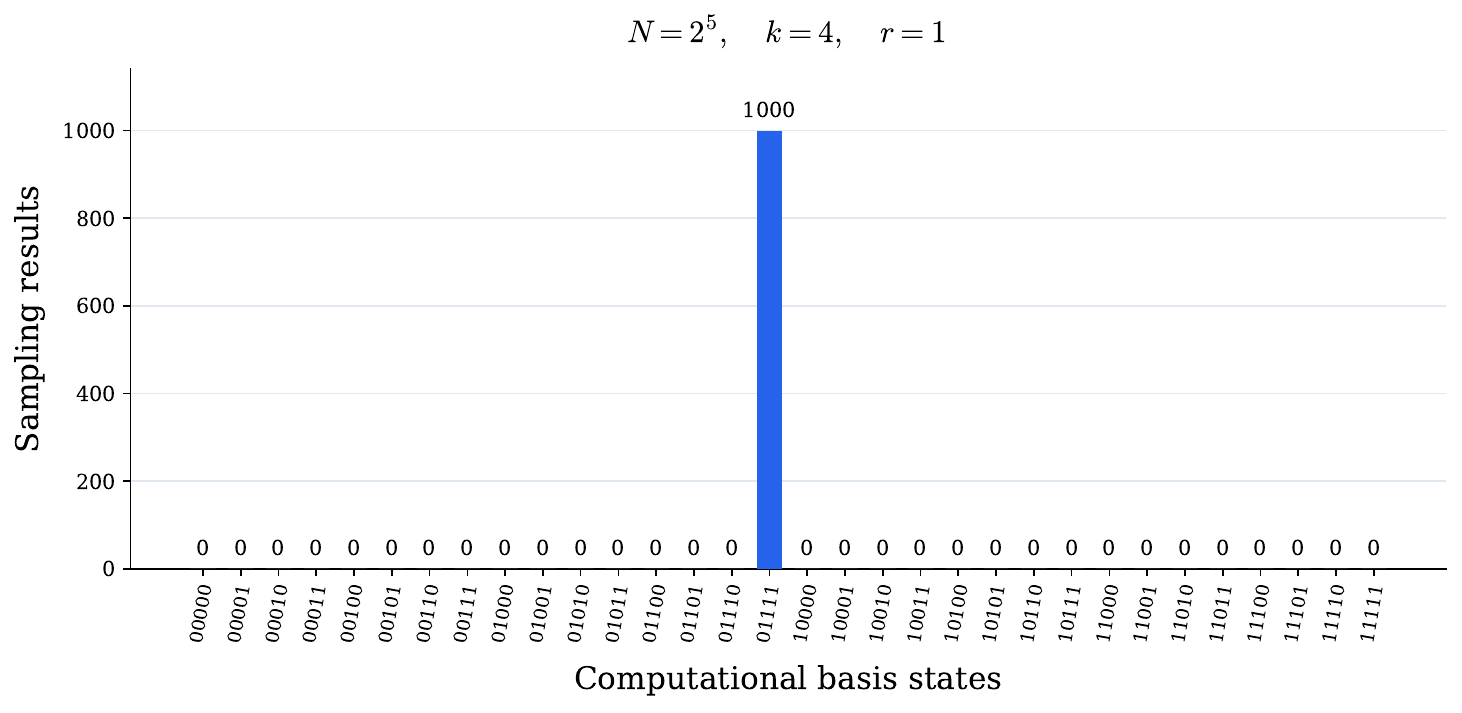}
 	\includegraphics[width=0.49\textwidth]
 	{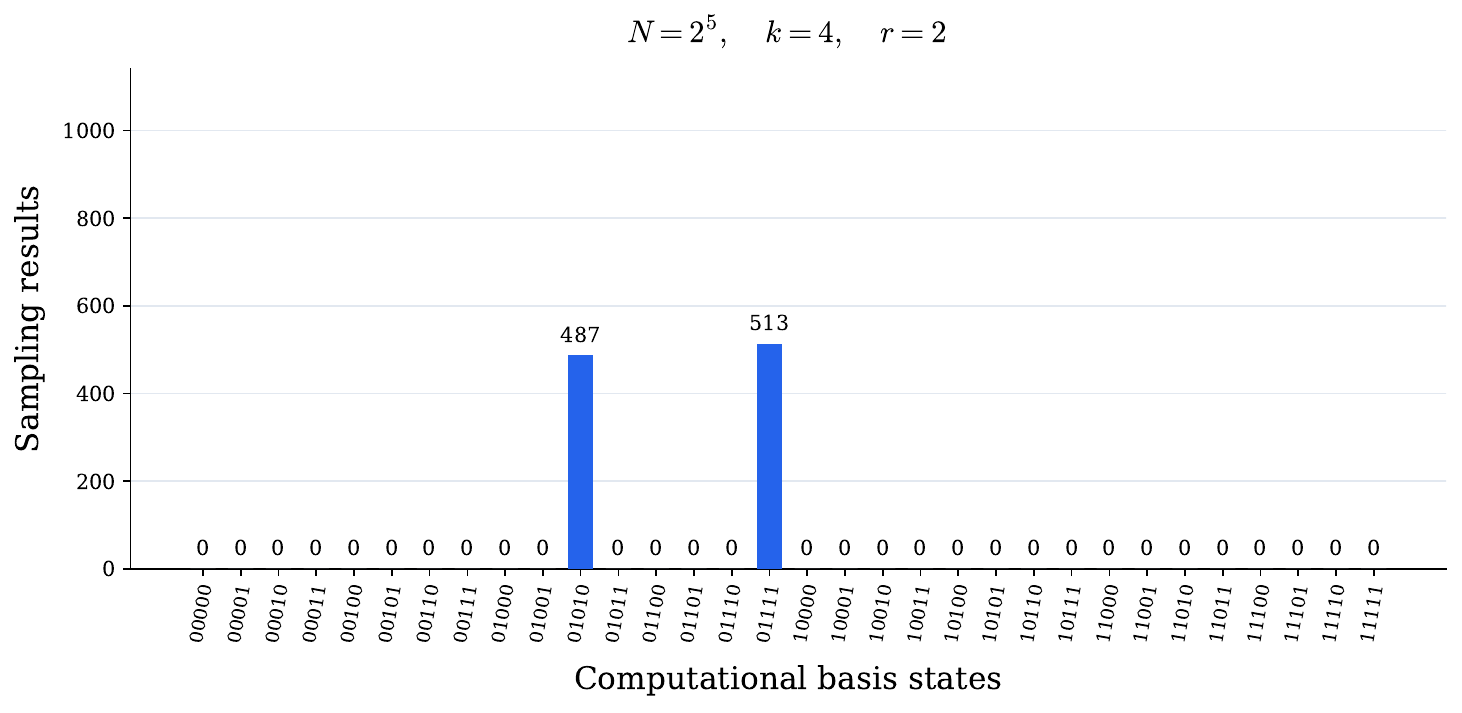}
 	\includegraphics[width=0.49\textwidth]
 	{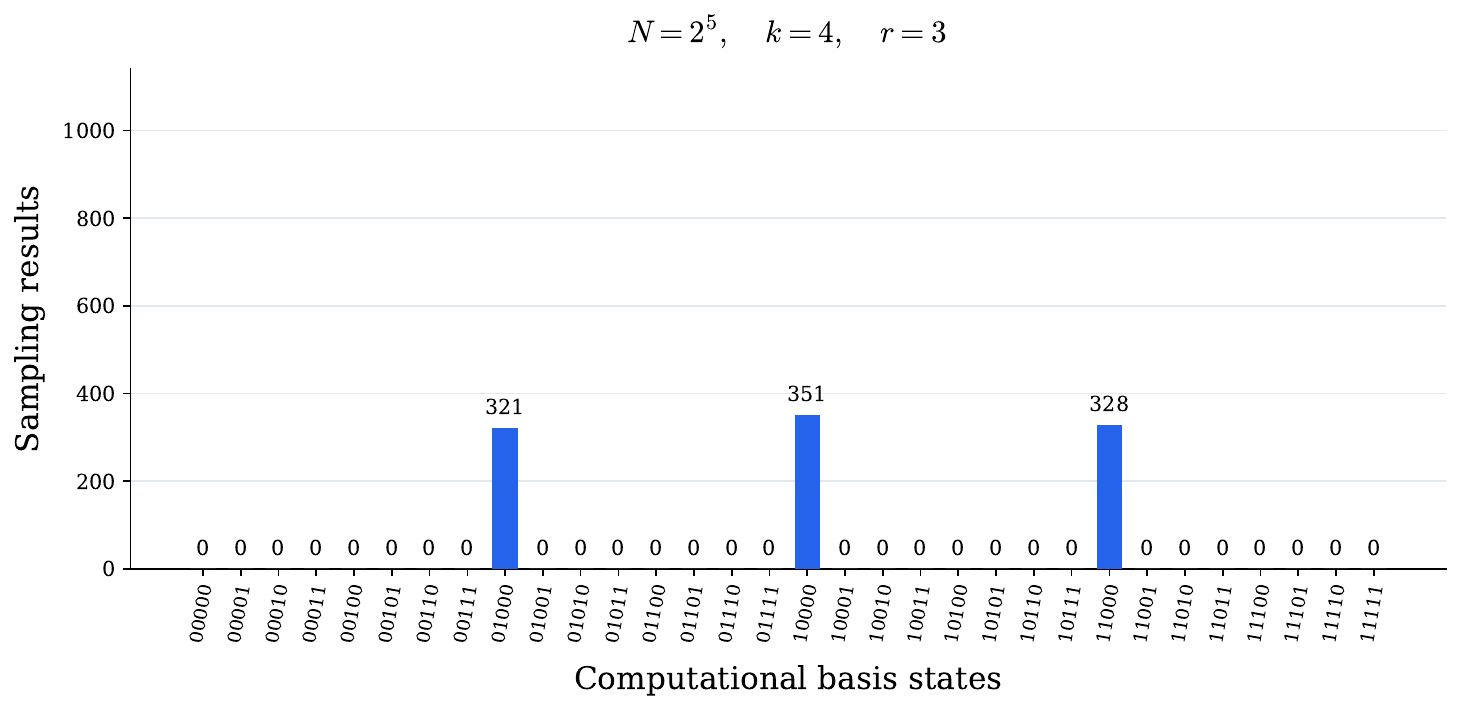}
 	\includegraphics[width=0.49\textwidth]
 	{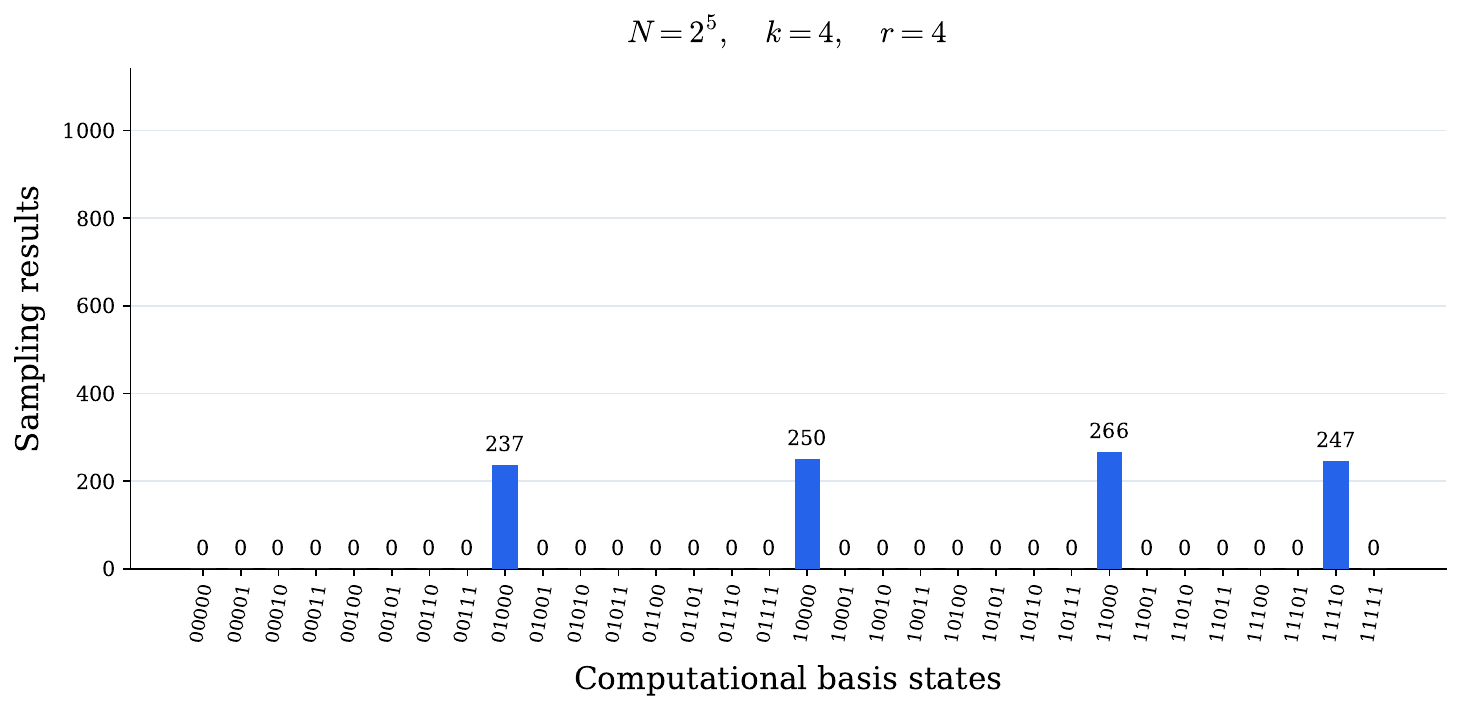}
 	\caption{
 		Experimental results for the exact search of the intersection of four sets for $r=1$, $2$, $3$, and $4$.
 	}
 	\label{fig:k4-results}
\end{figure*}

Fig.~\ref{fig:k4-results} shows the sampling results for $k=4$ with $r=1, 2, 3,$ and $4$. For $r=1, 2, 3,$ and $4$, the target states correspond to $\{|01111\rangle\}$, $\{|01010\rangle, |01111\rangle\}$, $\{|01000\rangle, |10000\rangle, |11000\rangle\}$, and $\{|01000\rangle, |10000\rangle, |11000\rangle, |11110\rangle\}$, respectively. The numbers of iterations $L$ are 4, 3, 3, and 2, respectively.The same result as above can also be obtained from Fig.~\ref{fig:k4-results}.
 
\section{Conclusion}
\label{sec:5}

In this work, we propose an algorithm for exactly finding the intersection of multiple sets. We derive and prove a phase-matching condition for finding the intersection with certainty. Furthermore, we present the quantum circuits for exact intersection search. Numerical simulations of the quantum circuits were conducted to validate the theoretical results and illustrate the effectiveness of the proposed method under different parameter settings. Our algorithm may be appreciated when the dimension is not large and certainty is important, as well as in situations where the preparation of the initial state and changes of experimental settings during the computation process are difficult.

\subsection*{Acknowledgement}

This research is supported by Zhejiang Provincial Natural Science
Foundation of China under Grant No. LZ24A050005.

\subsection*{Data availability statement}

The data and code that support the findings of this study are available from the corresponding author upon reasonable request.

\subsection*{Declaration on the use of generative AI}

The authors designed the study, carried out the mathematical analysis, and prepared the manuscript. Generative-AI tools were used for language refinement. They were also used to assist in writing the experimental code. All mathematical statements and experimental code were independently reviewed, tested, and verified by the authors, who take full responsibility for the content.


\end{document}